# Suppression of Multilayer Graphene Patches during Graphene growth

Zheng Han[*], Amina Kimouche, Adrien Allain, Hadi Arjmandi-Tash, Antoine Reserbat-Plantey, Sébastien Pairis, Valérie Reita, Nedjma Bendiab, Johann Coraux, and Vincent Bouchiat

Institut NÉEL, CNRS & Université Joseph Fourier, BP166, F-38042 Grenoble Cedex 9, France.

Keywords: chemical vapor deposition, single-layered graphene, pulsed growth, copper

Supporting Information Placeholder

**ABSTRACT:** By limiting the carbon segregation at the copper surface defects, a pulsed chemical vapor deposition method for single layer graphene growth is shown to inhibit the formation of few-layer regions, leading to a fully single-layered graphene homogeneous at the centimeter scale. Graphene field-effect devices obtained after transfer of pulsed grown graphene on oxidized silicon exhibit mobilities above 5000 $cm^2V^{-1}s^{-1}$.

Controlling the synthesis of graphene of good crystalline quality and its transfer onto an arbitrary substrate is a major stake in graphene research.[1-4] Many applications, like in photovoltaics,[5,6] transparent electrodes,[7,8] or in batch-production of nano-electronic devices,[9,10] require homogeneous graphene with continuous area well above 1 cm². Among emerging methods,[11] the chemical vapor deposition (CVD) of graphene on copper is an outstanding one, as it meets the two requirements of large size growth of good quality graphene and easy transfer onto arbitrary substrates.[12]

The low solubility of carbon in copper[13] confines the growth of graphene on the Cu surface, which becomes catalytically inactive once fully covered with graphene. In principle, this self-limits the growth to a single layer of graphene.[14] However, and to our knowledge, in all reports of CVD graphene on polycrystalline Cu, one could always observe the occurrence of patches of few-layer graphene,[15,16] having a typical area of a few µm² and covering up to ~ 10% of the total surface. These patches can be attributed to the carbon segregation which occurs preferentially at defect sites. Since graphene is known to exhibit completely different electronic properties with a slight change in its number of layers, it is paramount to achieve the production of a large-size fully single-layer graphene onto standard Cu foils excluding multilayers patches. The principle of the method we have optimized to reach that goal is based on increasing the relative exposure to hydrogen during the growth. This is achieved by alternating growth and reductive stages using successive exposures to hydrogen and hydrogen/carbon precursor mixture in an intermittent fashion.

The reductive action of hydrogen exposure at high temperature is known to efficiently etch graphene and to limit its growth. Accordingly, hexagonally shaped holes[17] and crystals[3] with straights edges can be engineered in graphene sheets. A hydrogen atmosphere is also expected to suppress carbon enrichment at high temperature at the defect sites in Cu, thus to allow to efficiently inhibit carbon segregation. This property suggests that, by properly adjusting the reductive action of hydrogen during CVD of graphene, one can open the route to: (i) the reduction of the density of defects through the control of the shape of the growing graphene islands, and (ii) to reduce the occurrence of multilayer patches through the suppression of carbon segregation from defect sites in Cu. Following that ideas, we present in the following a modification of the seminal copper-based growth technique,[12] which we will refer in the following as pulsed-CVD. We show that this procedure leads to large-size exclusively single-layer graphene, with mobility exceeding 5000 $cm^2V^{-1}s^{-1}$ after being transferred on a standard Si/SiO$_2$ wafer.

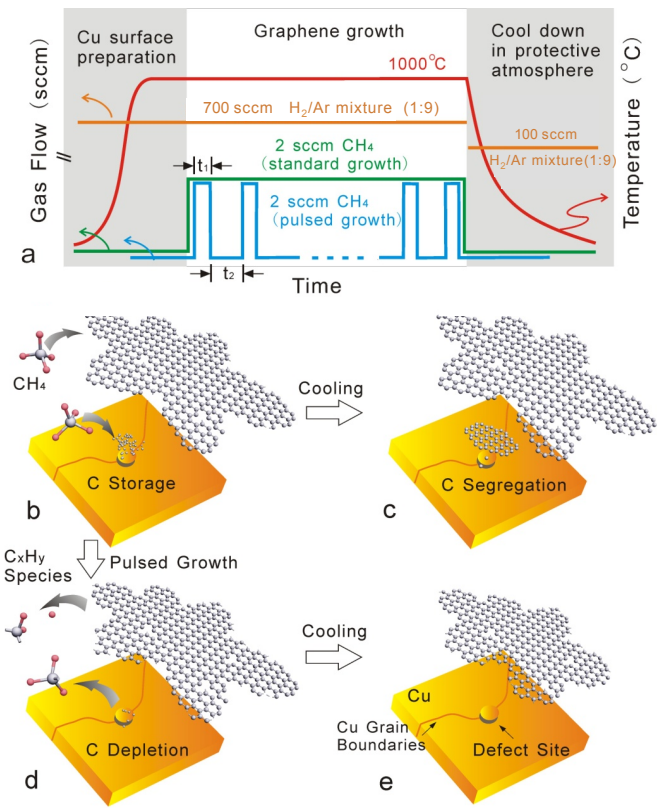

**Figure 1.** a) Process time flow showing a comparison the two methods for pulsed and standard growth. Detailed parameters can be found in the text. b)-e) Schematics showing the depleting of carbon at the defect sites thanks to the intermittent H$_2$ exposure, which prevents the formation of few-layer graphene patches underneath the first layer. Distances between graphene and Cu and substrate and graphene lattice are not to scale.

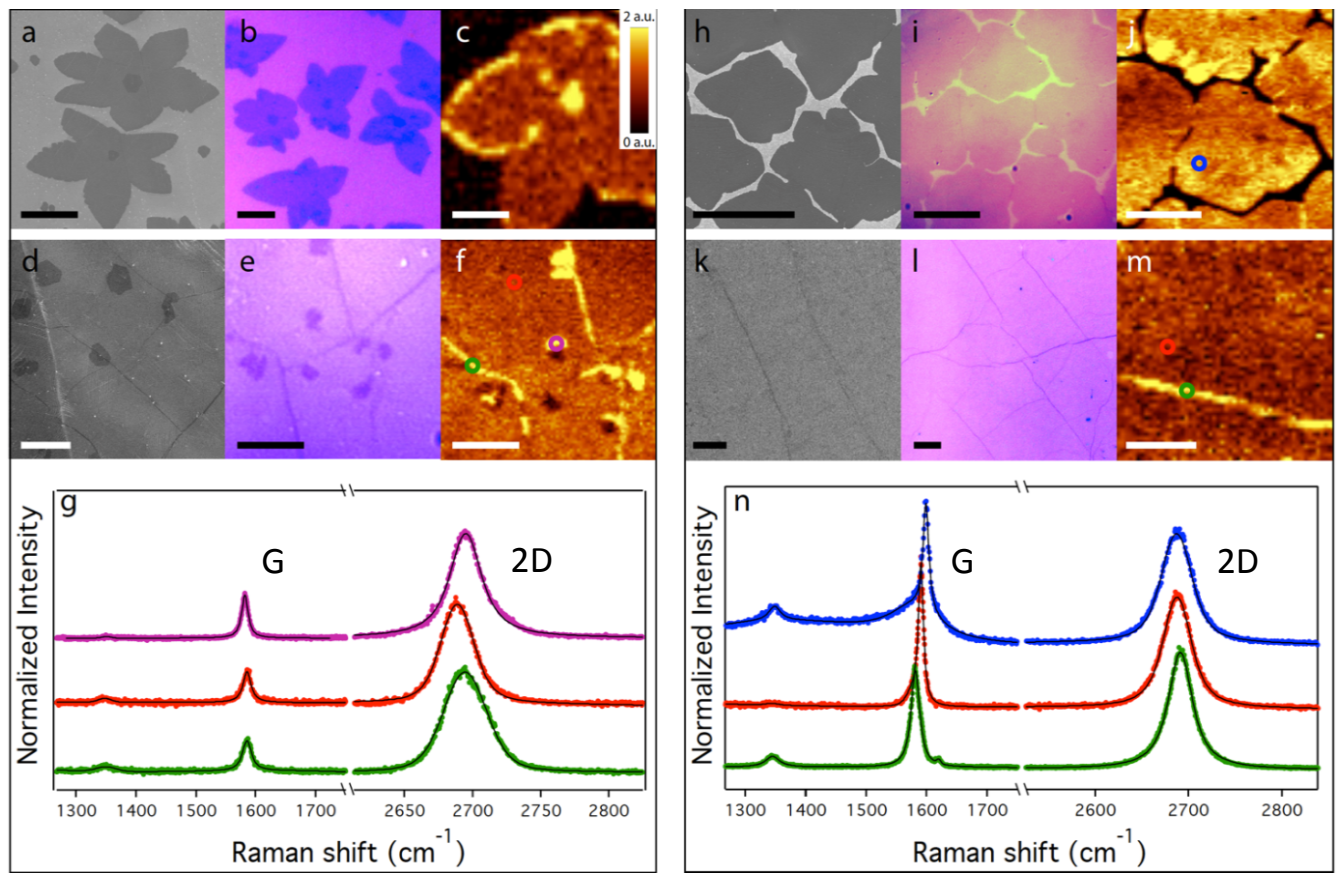

**Figure 2.** SEM, optical microscopy, Raman 2D intensity mapping, and Raman single spectra of graphene, prepared by the standard growth (Left panels) and pulsed growth methods (Right Panel) a-c, and d-f are graphene islands formed after 20 min, and full layer of graphene formed after 2 hour of standard growth, respectively. Single spectra in Fig. 2g are taken from different places in Fig. 2f (indicated by color). Right panel (pulsed growth): h-j, and k-m are graphene islands formed after 20 min (120 cycles), and full layer after 50 min (300 cycles) of pulsed growth, respectively. Fig. 2e and f are taken at the same place. Single spectra in Fig. 2n are taken from different places in Fig. 2j and 2f (indicated by color). For Raman single spectra, black solid lines are fitted by lorentzian. All growths are made under flow rates of 2 sccm $CH_4$ and 700 sccm $H_2$/Ar at 1 mbar pressure. During the pulsed growth process, $t_1 = 10$ s duration and $t_2 = 60$ s period are employed.

Unlike the standard conventional CVD process for graphene growth on catalytic surfaces, which is usually conducted at a fixed temperature and a constant flow of a carbon-source/$H_2$/Ar mixture for single continuous growth stage, our pulsed-CVD technique consists in a sequence of short (few seconds) growth time slots during which gaseous carbon precursor (methane) is introduced, separated by intervals free of carbon source keeping a constant $H_2$/Ar (1:9) mixture flow (Fig. 1a). In this work, we focus on the low-pressure (1 mbar) growth for the pulsed-CVD process.

First, as a reference, a standard conventional CVD process is carried out following the recipe described in Ref. 11. Scanning electron micrographs (SEM) of a typical sample after growth (Fig. 2a and Fig. S1) shows the graphene-covered Cu surface obtained after a 20 min growth time with a continuous flow of 2 standard cubic centimeters per minute (sccm) $CH_4$ diluted in 700 sccm $H_2$/Ar mixture. One clearly identifies graphene islands before they merge to form a continuous monolayer. Two island geometries can be found: lobe-shaped graphene "flowers" with diffusion-limited growth characteristics[18] and smaller hexagonally shaped ones (either centered underneath the flower-like graphene,[3,16,19,20] or isolated) (Fig. 2a). A rough analysis of the interactions between graphene nuclei and the Cu substrate can be found in Fig. S1. It is known that multilayer and inhomogeneous graphene results from enhanced carbon segregation which is favored at defects such as grain or stacking domain boundaries (e.g. screw dislocations or grain boundaries, see Fig. 1b-c and Fig. S2)[20,21] Therefore, one reasonable explanation could be that the larger islands are the result of surface growth from an adatom carbon gas at $T_g$, and that the smaller islands develop upon cooling down after growth, via segregation of carbon atoms stored at $T_g$ at Cu defects extending out of the surface plane (Fig. 1b-c). One way to suppress this defects-induced multi-layer islands segregation is to grow graphene on single crystal epitaxial films,[21] but the application of this technique is limited as the industrial-level large-scale single-crystal metal thin films would be tedious and too costly to fabricate.

Prolonged (2 hour) standard growth results in full coverage of the Cu foil with graphene, with prominently single layer regions but a few percent of the surface with two or more layers (Figs. 2d-f), which is in agreement with many previous works.[15,16] Optical images of both partial and full coverage of graphene transferred onto a $SiO_2$/Si wafer, shown in Figs. 2b and 2e, reveal the same features as in the corresponding SEM images. The Raman spectroscopy reveals Lorentzian-shaped G bands whose intensity is twice larger at the central point (Fig. 2g), suggesting non-Bernal stacking of bilayer graphene[22-24] at the center of the graphene islands. Optical contrast analysis (Fig. S3) further confirms this view. Earlier report[3] indicated that non-Bernal stacking of three or even more layers can be found on Cu by CVD. Except for the multi-layer regions, there are also lines showing higher 2D intensity, which are attributed to wrinkles created by the mismatch of thermal expansion coefficients of Cu and graphene, and have no direct connection with the graphene grain boundaries.[14,25]

Interestingly, after transfer onto a $SiO_2$/Si wafer, the sample presents spatial variations of the 2D band position of about 10 cm$^{-1}$. These variations follow the step edges of the Cu surface, which are observed with the secondary electron detector in SEM (Fig. S4). This may indicate a built-in stress in CVD graphene, which could be preserved even after transfer.

The outcome of a pulsed growth with 2 sccm $CH_4$, $t_1$ = 10 s and $t_2$ = 60 with 120 pulses, which corresponds to the same carbon dose as for Figs. 2a-c, is shown Figs. 2h-j. The main differences as compared to standard growth are: (i) more homogeneous size distribution of graphene islands, (ii) the absence of small hexagonal islands out of the large graphene islands.

Strikingly, unlike in any other reports on CVD graphene on Cu foil so far, no two-layer regions are observed on the islands, as shown in the SEM image in Fig. 2h and the optical image (Fig. 2i) of the same sample transferred onto $SiO_2$/Si wafer. This is further confirmed by Raman spectroscopy mapping, shown in Fig. 2j, where homogeneous 2D peak intensity is found over the whole island surface. Raman spectroscopy analysis of this homogeneous region indicates the presence of single-layer graphene as suggested by the typical spectra presented in Fig. 2n (blue point). Besides the narrow size distribution, the edges of lobe-shaped graphene island in pulsed growth are smoother, as can be seen by comparing Fig. 2a and 2h. fractality analysis plotting the average ratio of island perimeter as a function of its area shows that graphene islands are growing by pulsed CVD into a less dendritic form, namely, with smoother edges, than those grown by the standard method (Fig. S5, suppl. info).

The above results can be understood under the reasonable assumption that the underneath second layer regions are formed by carbon segregation upon cooling down. In pulsed growth, except during $CH_4$ pulses, the Cu defects are continuously depleted from carbon due to the reducing $H_2$ atmosphere, so that cooling down is not accompanied by noticeable carbon surface segregation, as shown in the schematic picture in Fig. 1b-e. Only surface growth from a carbon adatom gas is active in this case. As a result, the size distribution of graphene islands does not exhibit two modes as for standard growth, *i.e.*, the island shape is better controlled with pulsed growth.

When the number of pulses is increased from 120 to 300 pulses, with 2 sccm $CH_4$, $t_1$ = 10 s and $t_2$ = 60 s, full coverage of purely single layer graphene is obtained, as can be seen in both the SEM image of the sample on Cu (Fig. 2k), and the optical image after transfer (Fig. 2l). The fact that 2D peak Raman intensity mapping (Fig. 2m) is very homogeneous except at the thermal-expansion-induced wrinkles proves that there is no multilayer region. We will address the issue of unfolding these wrinkles in a future study. The Lorentzian profile, the position and the FWHM of the Raman 2D band are similar before and after coalescence of the graphene islands (Fig. 2n, red). The absence of any multilayer patch has been carefully checked on all samples and has been demonstrated on a millimeter square sample transferred onto $SiO_2$/Si wafer (Fig. S6). A comparison of optical images between standard and pulsed of graphene after transfer is given in Fig. 3. The coverage with single-layer graphene in our samples exceeds 99.9%. Regions without graphene (less than 0.1% of the surface, as marked also by circle region in Fig. S6) presumably develop from defects in graphene upon the effect of hydrogen etching due to Cu contaminations (Fig. S7).[3,17] In this study, the methane pulse duration $t_1$, and the interval between pulses $t_2$ have been varied over a wide range in order to reach optimal control over the density, shape and coverage rate of graphene islands. Once it is detuned far away from their optimal ratios, one observes either a very low graphene coverage (sometimes even no carbon deposits) or on the other side, the appearance of multi-layer patches. Both cases can be attributed from an imbalance between etching and carbon deposition. The qualitative diagram presented in Fig. S8 points out the parameter-space region for fully single-layer growth. To our knowledge, it is the first time a large-scale fully single-layer graphene has been achieved by CVD on Cu foils.

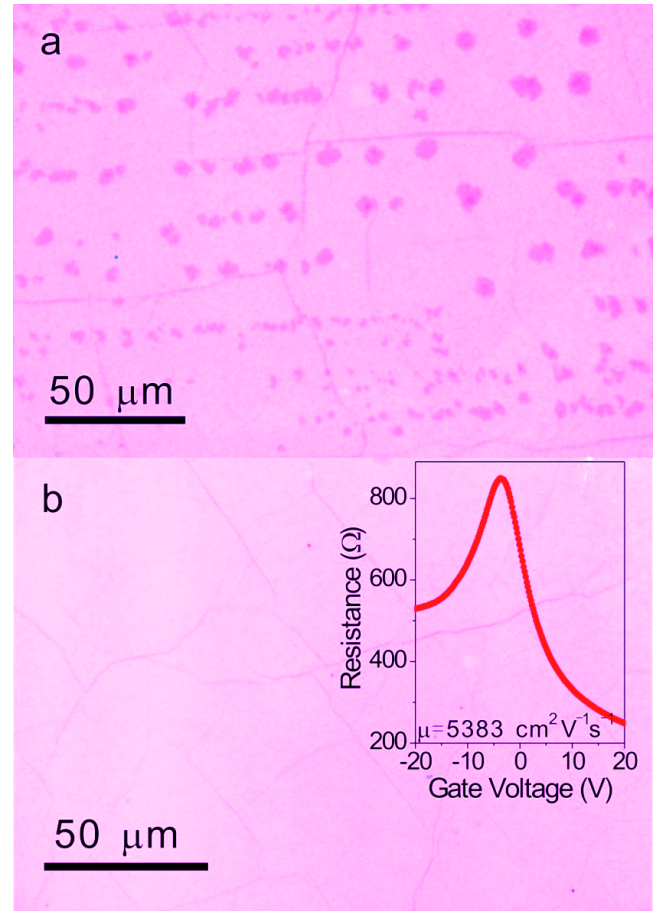

**Figure 3.** Optical micrographs of graphene after transfert on Si/$SiO_2$ substrate. a) Typical standard CVD growth. showing multi-layer patches b) Typical Pulsed CVD method. Inset in b) is a field effect curve of a device obtained by etching and connecting a micron-sized ribbon, of sample depicted in b, showing an electron mobility of 5383 $cm^2V^{-1}s^{-1}$.

Compared to exfoliated graphene, graphene prepared by CVD always usually exhibits lower carrier mobility,[11] which can be ascribed to both its polycrystalline nature and defects induced by the transfer process. Measuring the samples mobility is a convenient way to benchmark the quality of graphene in the view of transport measurements. To the exception of very high values reaching almost 20 000 $cm^2V^{-1}s^{-1}$, which were reported by R. S. Ruoff and co-workers,[26] mobilities in CVD-derived graphene are usually at best a few 1000 $cm^2V^{-1}s^{-1}$. For example, with a modified clean transfer method, mobility of 1000-1400 $cm^2V^{-1}s^{-1}$ can be obtained with a narrow distribution.[27] Recently, values as high as 4000-6000 $cm^2V^{-1}s^{-1}$ were reported in CVD graphene, which is dominantly single layer with second layer patches.[25] In the pulsed

CVD graphene case, the maximum mobility is found to be 5383 $cm^2V^{-1}s^{-1}$ (inset of Fig. 3b), which lies in the upper range of the values reported for CVD graphene samples.

As a conclusion, we proposed a new route towards the suppression of multi-layer inhomogeneities in CVD graphene, which consists in depleting carbon storage at defects by applying intermittent growth under continuous hydrogen exposure (pulsed growth). This scalable technique is a low-cost alternative to the use of high-quality metallic surfaces (metal single crystals, epitaxial thin films).[21] The pulsed growth technique is probably compatible with many other graphene growth recipes, such as ambient pressure growth, and plasma enhanced growth. Graphene devices on a 285 nm $SiO_2$-capped Si wafer fabricated by this technique exhibits electronic mobility up to 5383 $cm^2V^{-1}s^{-1}$. The preparation of purely single-layer graphene has great potential for the realization of homogeneous graphene films for electronic or optical applications.

## ASSOCIATED CONTENT

**Supporting Information**. The contents of Supporting Information include the following: (1) Analysis of the crystallographic orientations on the graphene islands by standard CVD method, and the optical contrast analysis for them. (2) Schematic picture of the multilayer regions formed by carbon segregation during cooling down in the CVD process. (3) Built-in stress of CVD graphene after transfer onto $SiO_2$/Si wafer. (4) Hexagonal pits formed by hydrogen etching. (5) Optical images showing the multilayer patches in standard CVD growth and the large scale fully single-layer graphene formed by pulsed CVD method. (6) SEM image showing the re-appearance of multi-layer during a pulsed CVD method with a high flow rate of C feedstock.

## AUTHOR INFORMATION

**Corresponding Author**

* Vincent.bouchiat@grenoble.cnrs.fr

## ACKNOWLEDGMENT

The authors are indebted to Christophe Guttin, Julien Jarreau, and Richard Haettel for their support in setting up the CVD reactor. This work is partially supported by the French ANR contract ANR-2010-BLAN-SUPERGRAPH, Region Rhone-Alpes CIBLE program, the EU contract NMP3-SL-2010-246073 "GRENADA" and the Nanosciences Foundation of Grenoble.

## REFERENCES

(1) Rasool, H. I.; Song, E. B.; Allen, A. J.; Wassei, J. K.; Kaner, R. B.; Wang, K. L.; Weiller, B. H.; Gimzewski, J. K. *Nano Lett.* **2011**, *11*, 251

(2) Li, X.; Magnuson, C. W.; Venugopal, A.; Tromp, R. M.; Hannon, J. M.; Vogel, E. M.; Colombo, L.; Ruoff, R. S. *J. Am. Chem. Soc.* **2011**, *133*, 2816

(3) Vlassiouk, I.; Regmi, M.; Fulvio, P.; Dai, S.; Datskos, P.; Eres, G.; Smirnov, S. *ACS Nano* **2011**, *5*, 6069

(4) Yu, Q. K.; Jauregui, L. A.; Wu, W.; Colby, R.; Tian, J.; Su, Z. H.; Cao, H. L.; Liu, Z. Z.; Pandey, D.; Wei, D. G.; Chung, T. F.; Peng, P.; Guisinger, N. P.; Stach, E. A.; Bao, J. M.; Pei, S. S.; Chen, Y. P. *Nature Materials* **2011**, *10*, 443-449

(5) Gomez De Arco, L.; Zhang, Y.; Schlenker, C. W.; Ryu, K.; Thompson, M. E.; Zhou, C. *ACS Nano* **2010**, *4*, 2865

(6) Park, H.; Brown, P. R.; Bulović, V.; Kong, J. *Nano Lett.* **2012**, *12*, 133-140

(7) Bae, S.; Kim, H. K.; Lee, Y. B.; Xu, X. F.; Park, J. S.; Zheng, Y.; Balakrishnan, J.; Lei, T.; Kim, H. R.; Song, Y. I.; Kim, Y. J.; Kim, K. S.; Özyilmaz, B.; Ahn, J. H.; Hong, B. H.; Iijima, S. *Nat. Nanotech.* **2010**, *5*, 574-578

(8) Kim, K. S.; Zhao, Y.; Jang, H.; Lee, S. Y.; Kim, J. M.; Kim, K. S.; Ahn, J.-H.; Kim, P.; Choi, J.-Y.; Hong, B. H. *Nature* **2009**, *457*, 706

(9) Lin, Y. M.; Dimitrakopoulos, C.; Jenkins, K. A.; Farmer, D. B.; Chiu, H. Y.; Grill, A.; Avouris, Ph. *Science* **2010**, *327*, 662

(10) Avsar, A.; Yang, T. Y.; Bae, S.; Balakrishnan, J.; Volmer, F.; Jaiswal, M.; Yi, Z.; Ali, S. R.; Güntherodt, B.; Hong, B. H.; Beschoten, B.; Özyilmaz, B. *Nano Lett.* **2011**, *11*, 2363

(11) Cooper, D. R.; D'Anjou, B.; Ghattamaneni, N.; Harack, B.; Hilke, M.; Horth, A.; Majlis, N.; Massicotte, M.; Vandsburger, L.; Whiteway, E.; Yu, V. *arXiv*: 1110.6557v1

(12) Li, X.; Cai, W.; An, J.; Kim, S.; Nah, J.; Yang, D.; Piner, R.; Velamakanni, A.; Jung, I.; Tutuc, E.; Banerjee, S. K.; Colombo, L.; Ruoff, R. S. *Science* **2009**, *324*, 5932

(13) McLellan, R. B. *Scripta Met.* **1969**, *3*, 389

(14) Li, X.; Cai, W. W.; Colombo, L.; Ruoff, R. S. *Nano Letter* **2009**, *9*, 4268-4272

(15) Yan, K.; Peng, H. L.; Zhou, H.; Li, H.; Liu, Z. F. *Nano Letter* **2011**, *11*, 1106-1110

(16) Robertson, A. W.; Warner, J. H. *Nano Letter* **2011**, *11*, 1182-1189

(17) Zhang, Y.; Li, Z.; Kim, P.; Zhang, L. Y.; Zhou, C. W. *ACSnano* **2012**, *6*, 126-132

(18) Nie, S.; Wofford, J. M.; Bartlet, N. C.; Dubon, O. D.; McCarty, K. F. *Phys. Rev. B*, **2011**, *84*, 155425

(20) Han, G. H.; Gunes, F.; Bae, J. J.; Kim, E. S.; Chae, S. J.; Shin, H.-J.; Choi, J.-Y.; Pribat, D.; Lee, Y. H. *NanoLett.* **2011**, *11*, 4144-4148

(21) Yoshii, S.; Nozawa, K.; Toyoda, K.; Matsukawa, N.; Odagawa, A.; Tsujimura, A. *Nano Lett.* **2011**, *11*, 2628-2633

(22) Camara, N.; Huntzinger, J.-R.; Rius, G.; Tiberj, A.; Mestres, N.; Pérez-Murano, F.; Godignon, P.; Camassel, J. *Phys. Rev. B* **2009**, *80*, 125410

(23) Tiberj, A.; Camara, N.; Godignon, P.; Camassel, J. *Nanoscale Research Letters.* **2011**, *6*, 478

(24) Poncharal, P.; Ayari, A.; Michel, T.; Sauvajol, J.-L. *Phys. Rev. B* **2008**, *78*, 113407

(25) Ni, G. X.; Zheng, Y.; Bae, S.; Kim, H. R.; Pachoud, A.; Kim, Y. S.; Tan, C. L.; Im, D.; Ahn, J. H.; Hong, B. H.; Özyilmaz, B. *ACSnano* DOI: 10.1021/nn203775x

(26) Li, X.; Magnuson C. W.; Venugopal, A.; An, J.; Suk, J. W.; Han, B. Y.; Borysiak, M.; Cai, W. W.; Velamakanni, A.; Zhu, Y. W.; Fu, L. F.; Vogel, E. M.; Voelkl, E.; Colombo, L.; Ruoff, R. S. *Nano Lett.* **2010**, *10*, 4328-4334

(27) Liang, X. L.; Sperling, B. A.; Calizo, I.; Cheng, G. J.; Hacker, C. A.; Zhang, Q.; Obweng, Y.; Yan, K.; Peng, H. L.; Li, Q. L.; Zhu, X. X.; Yuan, Walker, A. R. H.; Liu, Z. F.; Peng, L. M.; Richter, C. A., *ACSnano* **2011**, *5*, 9144-9153